\documentclass[12pt,twocolumn]{iopart}
\usepackage{graphicx}
\usepackage{enumerate}
\newcommand{\beq}{\begin{equation}}
\newcommand{\eeq}{\end{equation}}
\newcommand{\bea}{\begin{eqnarray}}
\newcommand{\eea}{\end{eqnarray}}

\usepackage{iopams}
\usepackage{bm}
\begin{document}

\title[Antiferromagnetism in metals]{Antiferromagnetism in metals:\\ from the cuprate superconductors\\ to the
heavy fermion  materials}

\author{Subir Sachdev$^1$, Max A Metlitski$^2$ and Matthias Punk$^1$}

\address{$^1$ Department of Physics, Harvard University, Cambridge MA
02138, USA}
\address{$^2$ Kavli Institute for Theoretical Physics, University of California, Santa Barbara, CA 93106, USA}

\begin{abstract}
The critical theory of the onset of antiferromagnetism in metals, with concomitant Fermi surface reconstruction,
has recently been shown to be strongly coupled in two spatial dimensions. 
The onset of unconventional superconductivity 
near this critical point is reviewed: it involves a subtle interplay between the breakdown of fermionic quasiparticle excitations on the Fermi surface,
and the strong pairing glue provided by the antiferromagnetic fluctuations. The net result is a logarithm-squared enhancement of the pairing vertex for generic Fermi surfaces, 
with a universal dimensionless co-efficient {\em independent\/} of the strength of interactions, which is expected to lead
to superconductivity at the scale of the Fermi energy. 
We also discuss the possibility that the antiferromagnetic critical point can be replaced by an intermediate `fractionalized Fermi liquid' phase, in which there is Fermi surface reconstruction but no long-range antiferromagnetic order. We discuss the relevance
of this phase to the underdoped cuprates and the heavy-fermion materials.
\end{abstract}

\submitto{Journal of Physics: Condensed Matter,\\ SCES 2011 special issue}
\maketitle

\section{Introduction}
\label{sec:intro}

The study of quantum antiferromagnetism in metals is clearly of fundamental
importance to a variety of modern correlated electron materials, from the heavy-fermion
superconductors to the modern copper-based and iron-based high temperature superconductors \cite{keimer}.

As a prominent example, consider the electron-doped superconductor Nd$_{2-x}$Ce$_x$CuO$_4$.
Neutron scattering experiments \cite{greven07} demonstrated the onset of antiferromagnetic long-range order
in a metal at a doping $x \approx 0.14$, not too far from the dopings with the highest critical temperatures
for superconductivity. Early photoemission experiments \cite{armitage} also presented evidence for the reconstruction
of the Fermi surface near this doping. More recently, the reconstruction of the Fermi surface has been extensively
studied by quantum oscillation experiments in strong magnetic fields \cite{Kartsov1,Kartsov2}. And finally, 
transport experiments \cite{greene} have detected signatures of ``strange metal'' behavior near the onset of 
antiferromagnetism.

Similar physics also applies to the iron-based superconductors, as demonstrated by the example of 
BaFe$_2$(As$_{1-x}$Px)$_2$ \cite{matsuda}: here we find a quantum phase transition involving the
onset of antiferromagnetism, accompanied by high temperature superconductivity and strange metal behavior. 

The theory of the onset of antiferromagnetism in metals has been studied for many decades. It has recently
been established \cite{maxsdw1,maxsdw2} that the critical theory is strongly coupled in the physically 
important case of spatial dimension
$d=2$, with a breakdown of all the formal expansion methods of critical field theories. So accurate computations
which can be quantitatively compared with experiments are presently out of reach. Nevertheless, significant
qualitative insights have been gained, and here we will review the answers to two important questions:
\begin{enumerate}[(A)]
\item Does unconventional high temperature superconductivity appear near the antiferromagnetic critical point in metals ?
\item In the traditional Hartree-Fock 
theory of antiferromagnetism in metals, there is a single quantum critical point separating
the Fermi liquid with a ``large'' Fermi surface (a FL), from a Fermi liquid with antiferromagnetic order and a reconstructed Fermi surface of ``small pockets'' (an AFM-FL); note that the Fermi surface volumes obey the traditional Luttinger relation in both phases.
Can this critical point, under suitable conditions, be replaced by an intermediate non-Fermi liquid phase (or phases) ?
\end{enumerate}

The answer to question (A) will be presented in Section~\ref{sec:sc}. The proposal of $d$-wave-like pairing near
an antiferromagnetic quantum phase transition predates the discovery of the cuprate superconductors \cite{dwave1,dwave2,moriya}. 
At least in the 
weak-coupling limit, this proposal has been put on a solid footing \cite{raghu}. 
However, it has not been clear whether turning
up the strength of the interactions will lead to true higher temperature superconductivity. 
The stronger antiferromagnetic fluctuations can also degrade the integrity of the underlying fermionic quasiparticles, and this can compensate for any increase in the strength of the pairing glue \cite{msv}. Moreover, stronger interactions could lead to additional instabilities to other types of order, which can pre-empt superconductivity. In Section~\ref{sec:sc} we will review recent
computations \cite{maxsdw1} showing 
that high temperature superconductivity does indeed appear near the antiferromagnetic quantum critical point in two spatial dimensions, with the pairing glue dominating effects due to quasiparticle breakdown and to instabilities towards other orders.

Question (B) will be addressed in Section~\ref{sec:ffl}. We will review arguments that the single critical point can indeed
be replaced in appropriate conditions by an intermediate phase---the `fractionalized Fermi liquid' (FL*) \cite{ffl1,ffl2}. A complementary
review, with a more complete discussion of experiments and related theoretical work may be found in a recent paper by
M.~Vojta \cite{vojtarev}.
In the present context, the FL* phase
has its Fermi surface reconstructed into small pockets, but {\em without\/} antiferromagnetic order even at zero 
temperature \cite{qi,moon}; the absence of antiferromagnetic order implies that the Fermi surface volumes do {\em not\/}
obey the Luttinger relation in the FL* phase.
The traditional antiferromagnetic critical point is associated with two distinct changes in the ground state: the onset of antiferromagnetic order and the reconstruction of the Fermi surface. Section~\ref{sec:ffl} argues that these changes can be separated into two steps. Starting from the antiferromagnetic Fermi liquid with small pockets (AFM-FL), 
the first quantum transition involves the disappearance of antiferromagnetic order into a FL* phase \cite{qi,moon,ribhu,tarun}; however, the small pocket Fermi surfaces are retained in the FL* phase, even though they now violate the Luttinger relation. 
 The large Fermi surface appears only after 
one or more additional quantum transitions
lead eventually to a Fermi liquid with a large Fermi surface (FL). We will discuss applications of this exotic possibility of an intermediate FL* phase to the hole-doped cuprates. We also describe the appearance of the FL* phase in Kondo lattice models appropriate
to the heavy fermion compounds, where the FL* Fermi surfaces are associated with band structure of the conduction electrons.

\section{Superconductivity near the antiferromagnetic quantum critical point}
\label{sec:sc}

\begin{figure}
\centerline{\includegraphics[width=4.5in]{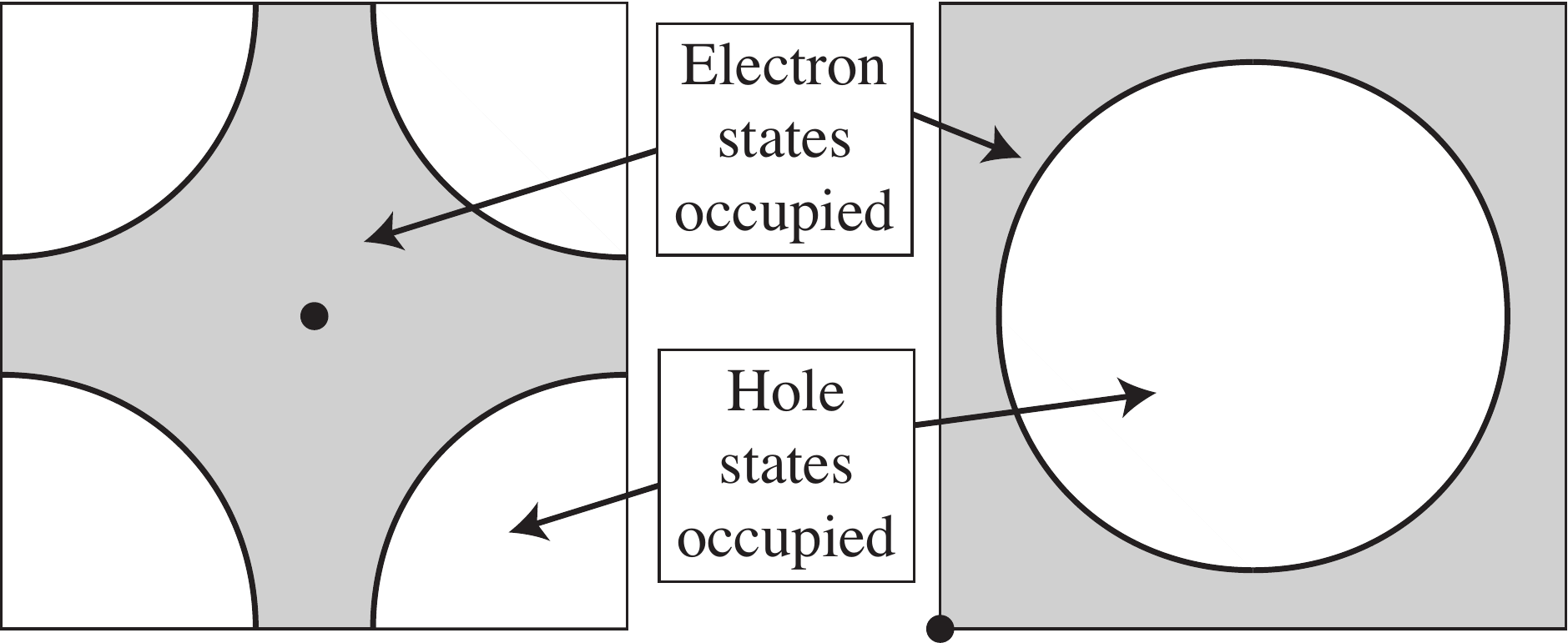}} \caption{Two views of the Fermi surface of the cuprate superconductors
(hole and electron doped) in the FL phase. 
The left panel has
the momentum ${\bf k}= (0,0)$ (the ``$\Gamma$ point'', denoted by the filled circle) 
in the center of the square Brillouin zone, while
the right panel has the $\Gamma$ point at the bottom-left edge. The momenta with both up and down electron states
occupied are shaded gray.
} \label{fig:cupratefs}
\end{figure}

\begin{figure}
\centering
 \includegraphics[width=3.5in]{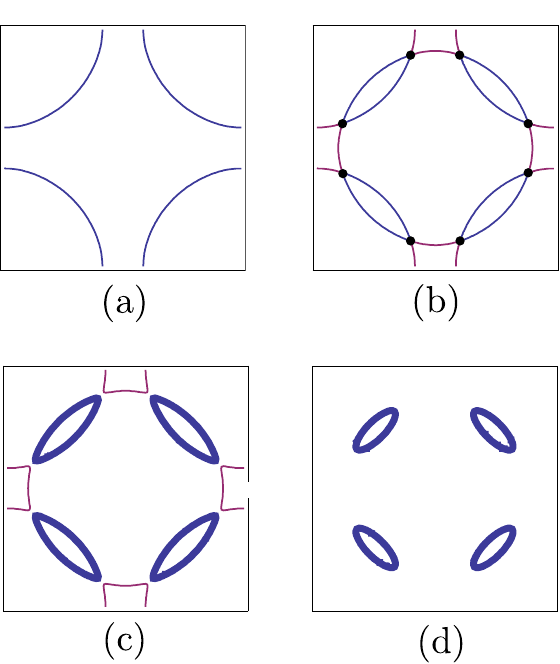}
 \caption{ The transformation of the Fermi surface of the cuprates
 by  antiferromagnetism \cite{SCS,CM}. The areas enclosed by the Fermi surfaces obey the Luttinger relation in all cases
 here. (a) Fermi surface without antiferromagnetic order (FL), as in Fig.~\ref{fig:cupratefs}
 (b) The original Fermi surface along with the Fermi surface shifted by wavevector ${\bf K} = (\pi, \pi)$. These 
 intersect at the hot spots shown by the filled circles. (c) With the onset of a non-zero spin 
 antiferromagnetic order with $\langle \varphi_\alpha \rangle \neq 0$ in the AFM-FL phase, 
 gaps open at the hot spots leading
 to electron (thin lines) and hole (thick lines) pockets. (d) With increasing $||\langle \varphi_\alpha \rangle ||$
 the electron pockets shrink to zero for the hole-doped case, leaving only hole pockets in a AFM-FL phase.
 In the electron-doped case, the hole pockets shrink to zero, leaving only electron pockets (this is not shown)
 in an AFM-FL phase.
 Finally, in the half-filled case, the electron and hole pockets shrink to zero simultaneously.}
 \label{fig:sssdw}
\end{figure}

In the familiar Hartree-Fock theory of antiferromagnetic (or spin density wave) ordering in a metal,
we begin with a Fermi liquid metal  (FL) associated with quasiparticles with a dispersion $\varepsilon_{\bf k}$: an example appropriate to the cuprates is shown in Fig.~\ref{fig:cupratefs}.
Then, we introduce spin density wave order at the wavevector ${\bf K} = (\pi, \pi)$, represented by the antiferromagnetic
order parameter $\varphi_\alpha$, with $\alpha = x, y, z$ the spin components. The electrons will undergo Bragg reflection
off this ordering, and so acquire a modified dispersion which will also change the Fermi surface. This is illustrated 
in Fig.~\ref{fig:sssdw}.
The antiferromagnetic order mixes electron states with momentum ${\bf k}$ and ${\bf k} + {\bf K}$, and so
in Fig.~\ref{fig:sssdw}b we plot the  Fermi surface along with the Fermi surface shifted by ${\bf K}$. There is Bragg reflection
of the zero-energy states on the Fermi surface only at the points where these Fermi surfaces intersect: these are the
``hot spots''. Gaps open at these hot spots, leading to a reconstruction of the Fermi surface into small pockets in the AFM-FL, as
shown in Fig.~\ref{fig:sssdw}c. 

\begin{figure}
\centering
 \includegraphics[width=4.5in]{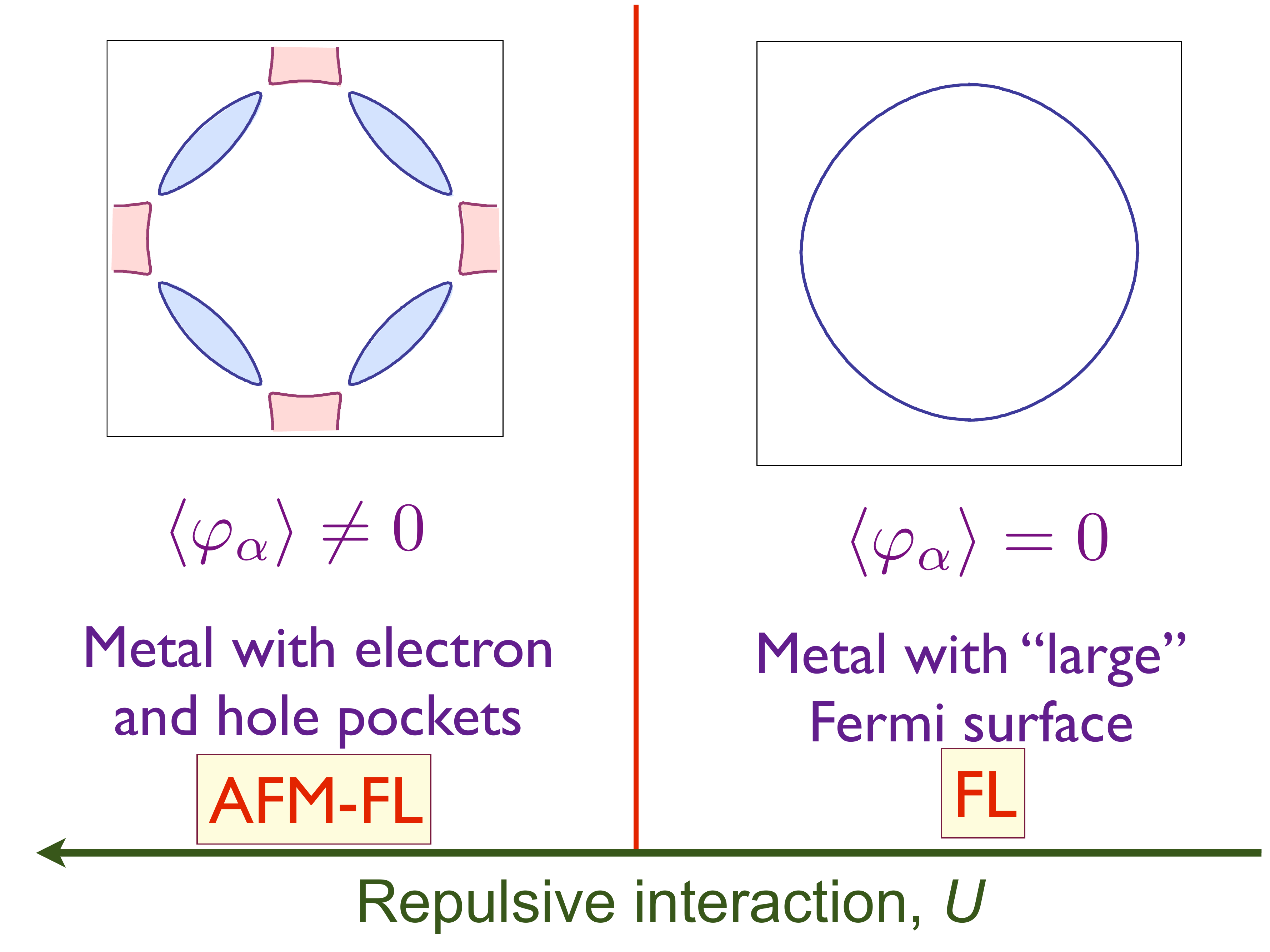}
 \caption{Zero temperature phases as a function of repulsive short-range interactions between the electrons.
 $\varphi_\alpha$ is the antiferromagnetic order parameter, and the Fermi surfaces 
 in the two phases are shown. The Fermi volumes obey the Luttinger relation in both phases. In the AFM-FL phase, the doubling
 of the unit cell by antiferromagnetic order ensures that the pocket Fermi surfaces are compatible with the Luttinger relation.}
 \label{fig:sdwphase1}
\end{figure}

We expect this transition involving onset of antiferromagnetism to occur as a function of increasing local repulsion 
between the electrons {\em e.g.\/} by turning up the strength of the on-site repulsion, $U$, in a Hubbard-like model.
This leads to the simple ground state phase diagram shown in Fig.~\ref{fig:sdwphase1}, with a quantum critical point at a critical interaction strength separating the phases with and without antiferromagnetic order.
The two phases have ``small'' and ``large'' Fermi surfaces respectively. The Luttinger relation counts the number of electrons modulo 2 per unit cell, and the doubling of the unit cell in the AFM-FL phase ensures that the Fermi surfaces in both phases enclose the traditional
Luttinger volume.

\begin{figure}
\centering
 \includegraphics[width=4.3in]{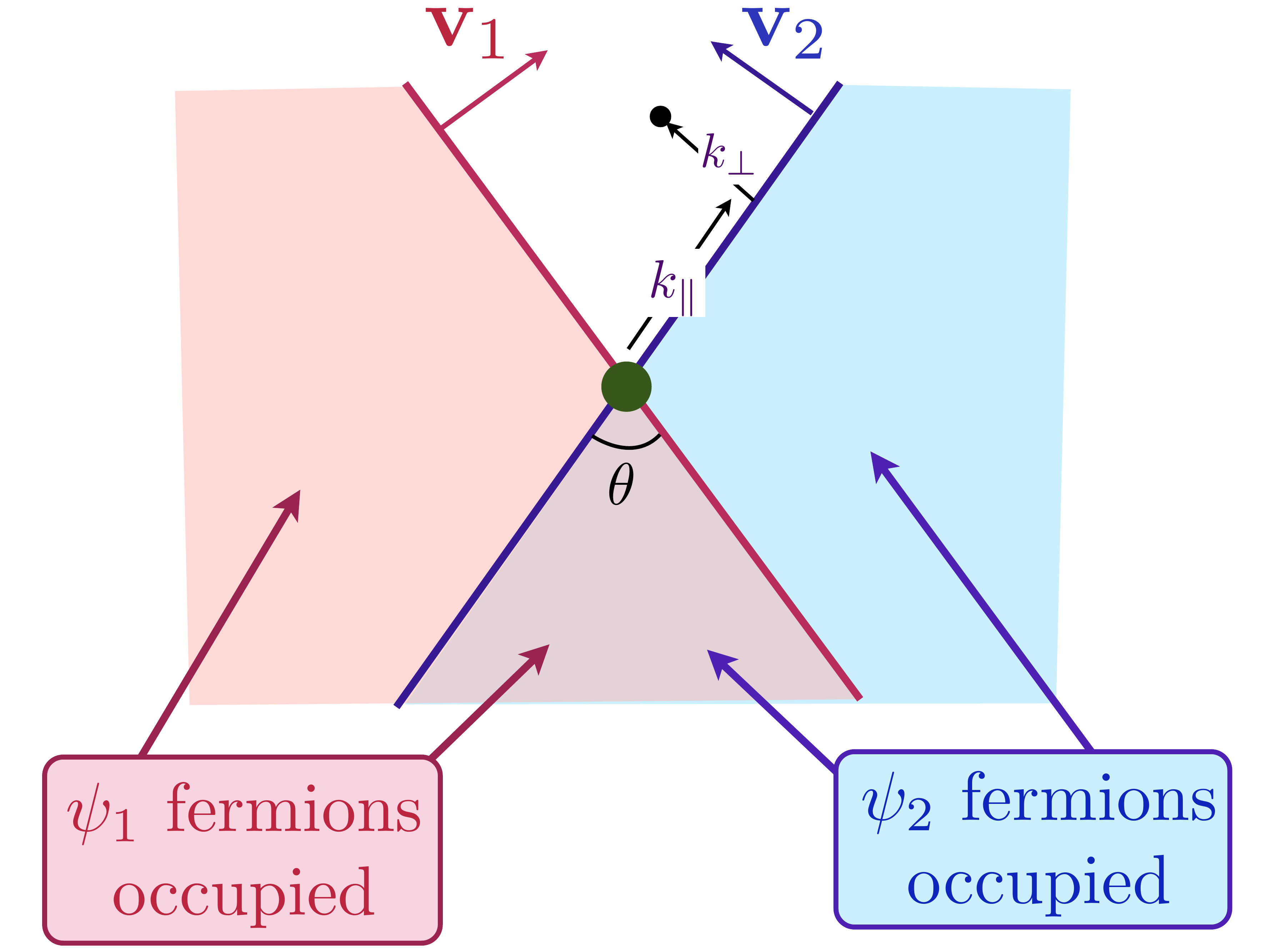}
 \caption{Fermi surfaces of $\psi_1$ and $\psi_2$ fermions and their respective Fermi velocities ${\bf v}_1$ 
 and ${\bf v}_2$ in the FL phase with $\langle\varphi_\alpha \rangle =0$. 
 The Fermi lines intersect at the hot spot, denoted by the filled circle at the origin.
 The vicinity of this hot spot is similar to any one of the hot spots in Fig.~\ref{fig:sssdw}b.}
 \label{fig:fermions}
\end{figure}

\begin{figure}
\centering
 \includegraphics[width=3.5in]{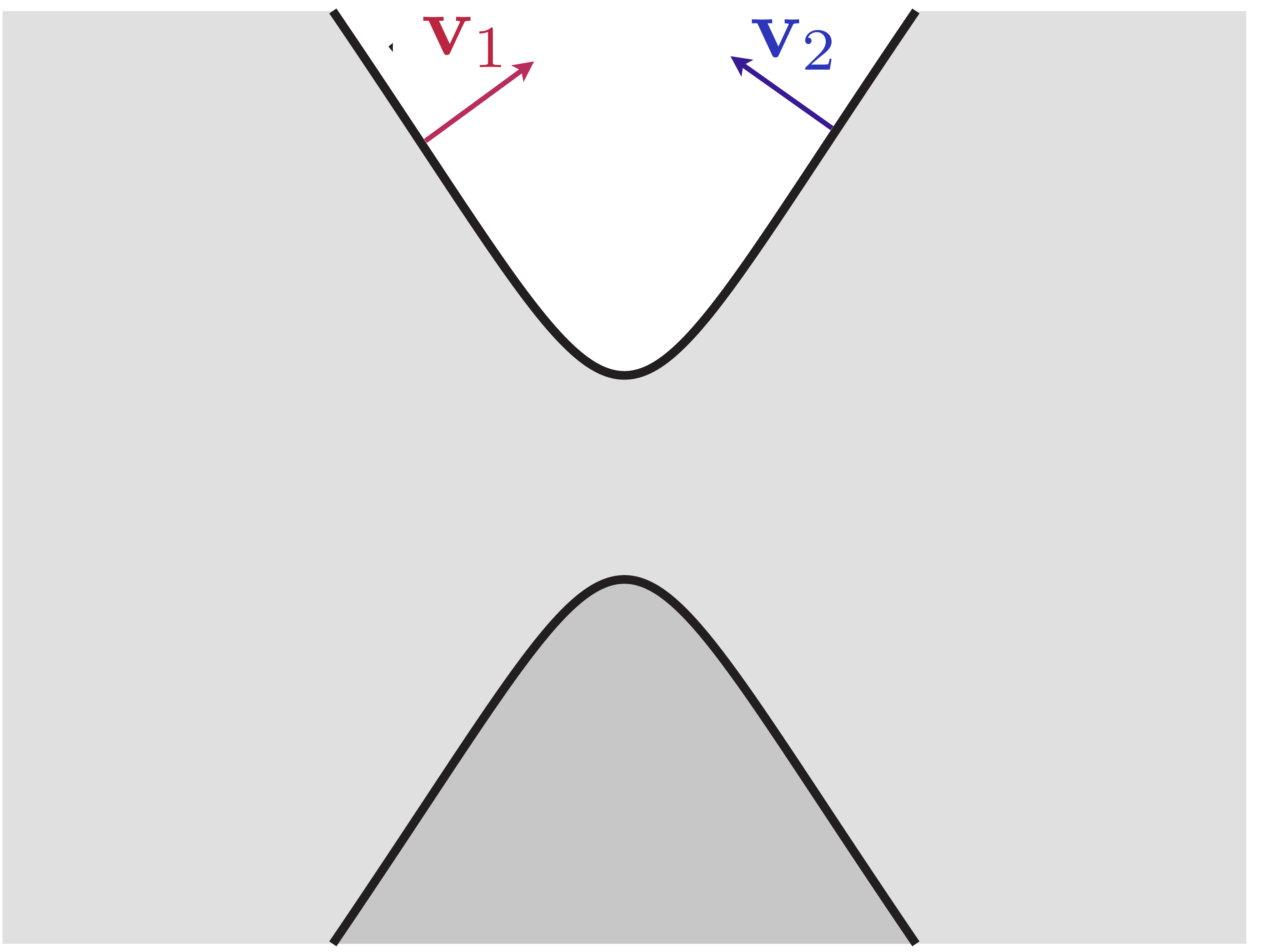}
 \caption{Deformation of the Fermi surfaces in Fig.~\ref{fig:fermions} in the AFM-FL phase 
 with $\langle \varphi_\alpha \rangle \neq 0$. A gap has opened up at the hot spot, and this leads to the Fermi surface reconstruction shown above, or equivalently, in Fig.\ref{fig:sssdw}c.}
 \label{fig:fermions_gap}
\end{figure}

We are interested here in the physics in the vicinity of this critical point. This is described by a 
universal low-energy theory \cite{AC} 
whose structure we now describe. The important low energy fermionic excitations lie in the vicinity of the hot spots; 
let us focus here on just one of the hot spots. There are two Fermi lines intersecting at the hot spots,
and we label the fermionic quasiparticles along these lines by $\psi_{1 a}$ and $\psi_{2 a}$ ($a = \uparrow, \downarrow$
is an electron spin label), as shown in Fig.~\ref{fig:fermions}. 
The momenta of both fermions will be measured with respect to the hot spot momentum ${\bf k}_h$. Then, these fermions are described by
the Lagrangian
\beq
\mathcal{S}_\psi = \int d \tau d^2 x \Biggl[ \psi_{1a}^\dagger \left( \frac{\partial}{\partial \tau} - i {\bf v}_1 \cdot {\bm\nabla}\right) \psi_{1a} + \psi_{2a}^\dagger  \left( \frac{\partial}{\partial \tau} - i {\bf v}_2 \cdot {\bm \nabla} \right) \psi_{2a} \Biggr],
\label{sdw2}
\eeq
where $\tau$ is imaginary time, ${\bf v}_1 = \left. {\bm \nabla}_k \varepsilon_{\bf k} \right |_{{\bf k}_h}$ is the Fermi velocity at ${\bf k}_h$, and similarly for ${\bf v}_2$. 
The critical theory is completed by coupling these fermions to the quantum fluctuations of the antiferromagnetic order parameter
$\varphi_\alpha$, described by
\beq
\mathcal{S}_{\psi\varphi} = \int d \tau d^2 x  \Biggl[ \frac{1}{2} ({\bm \nabla} \varphi_\alpha)^2 + \frac{r}{2} 
\varphi_\alpha^2
+ \frac{u}{4\!} \left( \varphi_\alpha^2 \right)^2 + \lambda \varphi_\alpha \sigma^{\alpha}_{ab} \left(
\psi_{1a}^\dagger \psi_{2b} + \psi_{2a}^\dagger \psi_{1b} \right) \Biggr], \label{sdw3}
\eeq
where $\sigma^\alpha$ are the Pauli matrices.
The first three terms in Eq.~(\ref{sdw3}) give the standard Landau-Ginzburg action representing the contribution of the high-energy electrons to the energy of the antiferromagnetic state. But the crucial term is the ``Yukawa'' coupling, $\lambda$ by which $\phi_\alpha$ scatters
a $\psi_1$ fermion into a $\psi_2$ fermion, and vice versa: in the original lattice co-ordinates, this is a process in which
the electron picks up a momentum close to ${\bf K}$ from the antiferromagnetic order parameter. This Yukawa coupling
can be obtained by a Hubbard-Stratanovich decoupling of the on-site interaction, in which case $\lambda^2 \sim U$.
In the AFM-FL phase where the expectation value $\langle \varphi_\alpha \rangle \neq 0$, the Yukawa coupling opens a gap of 
$2 \lambda ||\langle \varphi_\alpha \rangle ||$
at the hot spot, and this reconstructs the Fermi surfaces, as shown in Fig~\ref{fig:fermions_gap}.

As we are interested in the limit of large $U$, a bare perturbative analysis in powers of $\lambda$ is not expected to
be an acceptable strategy to analyze the critical theory.
A comprehensive study of the structure of the critical fluctuations of $\mathcal{S}_\psi + \mathcal{S}_{\psi\phi}$ has
been provided recently in Ref.~\cite{maxsdw1}. As noted earlier, there is a flow to strong coupling,
and so a complete understanding has not been achieved. Nevertheless, useful information can be obtained from a renormalized
two-loop analysis, and we summarize some of the basic results.

In the immediate vicinity of the hot spots, there is a complete breakdown of the fermionic quasiparticle excitations 
at the quantum critical point. The $\psi_1$ Green's function has the general structure \cite{AC,millis}
\beq
G_{\rm hot-spot}  \sim \frac{1}{\sqrt{i \omega} - {\bf v}_1 \cdot {\bf k}} \label{Ghot}
\eeq
where $\omega$ is a real frequency. So there is no quasiparticle pole, just a critical continuum of fermionic excitations
close to the hot spot. 
However, low energy quasiparticles are less strongly perturbed along the Fermi lines away from the hot spots. 
As denoted in Fig.~\ref{fig:fermions}, if we approach the Fermi line with $k_\perp \rightarrow 0$ at
a fixed $k_\parallel$, then the one-loop Fermi line Green's function does have a quasiparticle pole of the form \cite{maxsdw1}
\beq
G_{\rm Fermi-line} = \frac{Z (k_\parallel)}{\omega - v_F (k_\parallel) k_\perp} \label{Gline}
\eeq
where $Z(k_\parallel)$ is the quasi-particle residue, and $v_F (k_\parallel)$ is the renormalized Fermi velocity. 
The latter quantities were computed at the quantum critical point, and it was found that they both
vanish linearly with $k_\parallel$:
\beq
Z( k_\parallel)  \sim \lambda^{-2} k_\parallel \quad , \quad v_F (k_\parallel) \sim \lambda^{-2} k_\parallel. \label{kpar}
\eeq
For future convenience we have also indicated the dependence of $Z$, $v_F$ on the coupling constant $\lambda$. Thus, reassuringly, the quasiparticle residue does vanish as we approach the hot spot, which is consistent with 
(\ref{Ghot}). Indeed, we can deduce the structure of (\ref{kpar}) from (\ref{Ghot}), with the knowledge that $\omega$
scales as $k^2$ in the one-loop critical theory.

\subsection{Pairing instability}
\label{sec:pairing}

\begin{figure}
\centering
 \includegraphics[width=3.3in]{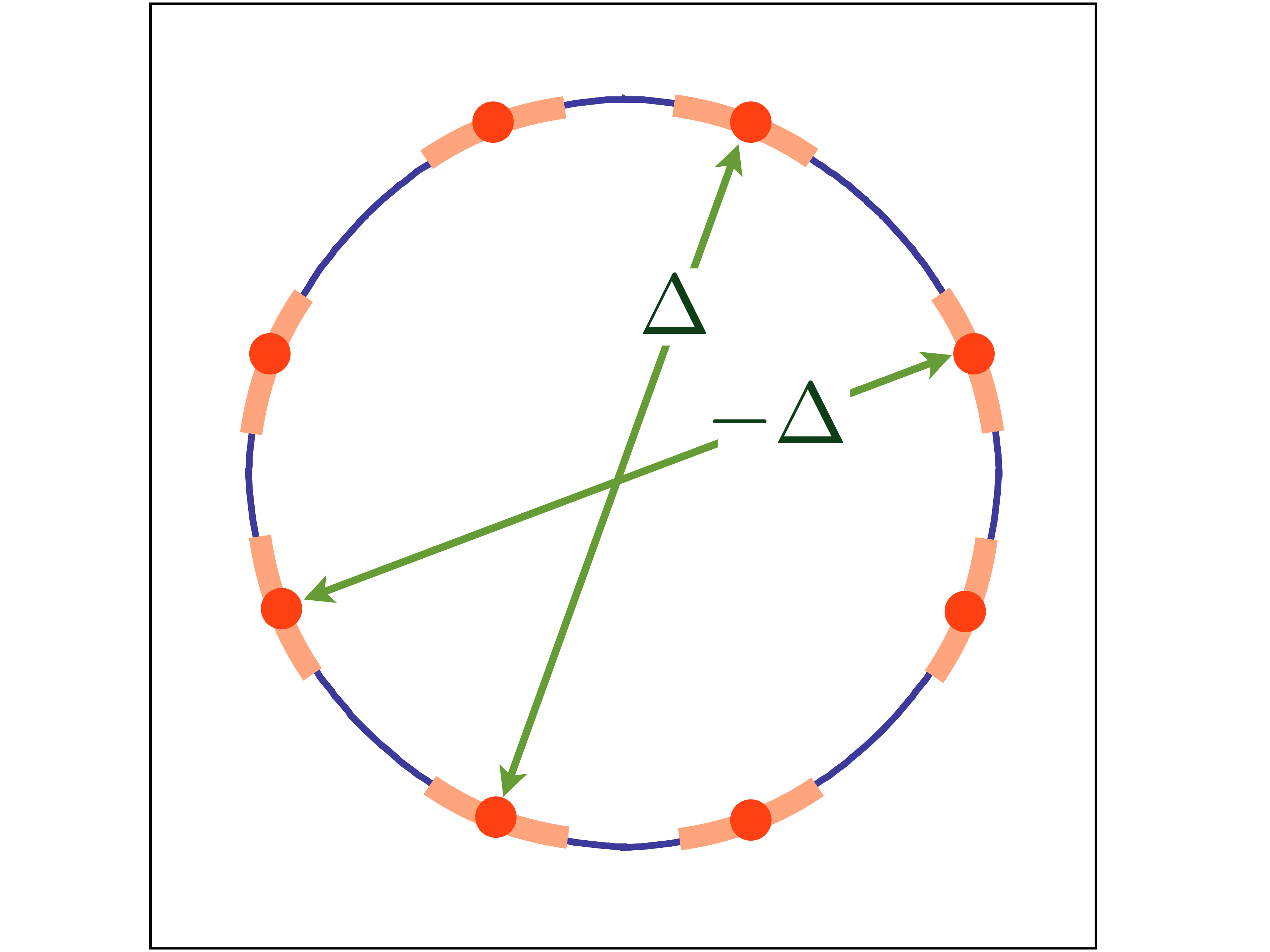}
 \caption{Pairing amplitude of the electrons near the hot spots. Hot spots separated by ${\bf K}$ have opposite signs
 for the pairing amplitude, leading to unconventional superconductivity.}
 \label{fig:pairing}
\end{figure}

We are now finally in a position to ask about the instability of the metal to pairing in the vicinity of the antiferromagnetic
quantum critical point. That such an instability is present, was pointed out already in an early study \cite{dwave1}. 
The instability was found to be towards unconventional superconductivity, in which the electrons whose momenta differ
by ${\bf K}$ have opposite signs for the pairing amplitude. This is illustrated in Fig.~\ref{fig:pairing} for the Fermi surface appropriate to the cuprates.

\begin{figure}
\centering
 \includegraphics[width=2.5in]{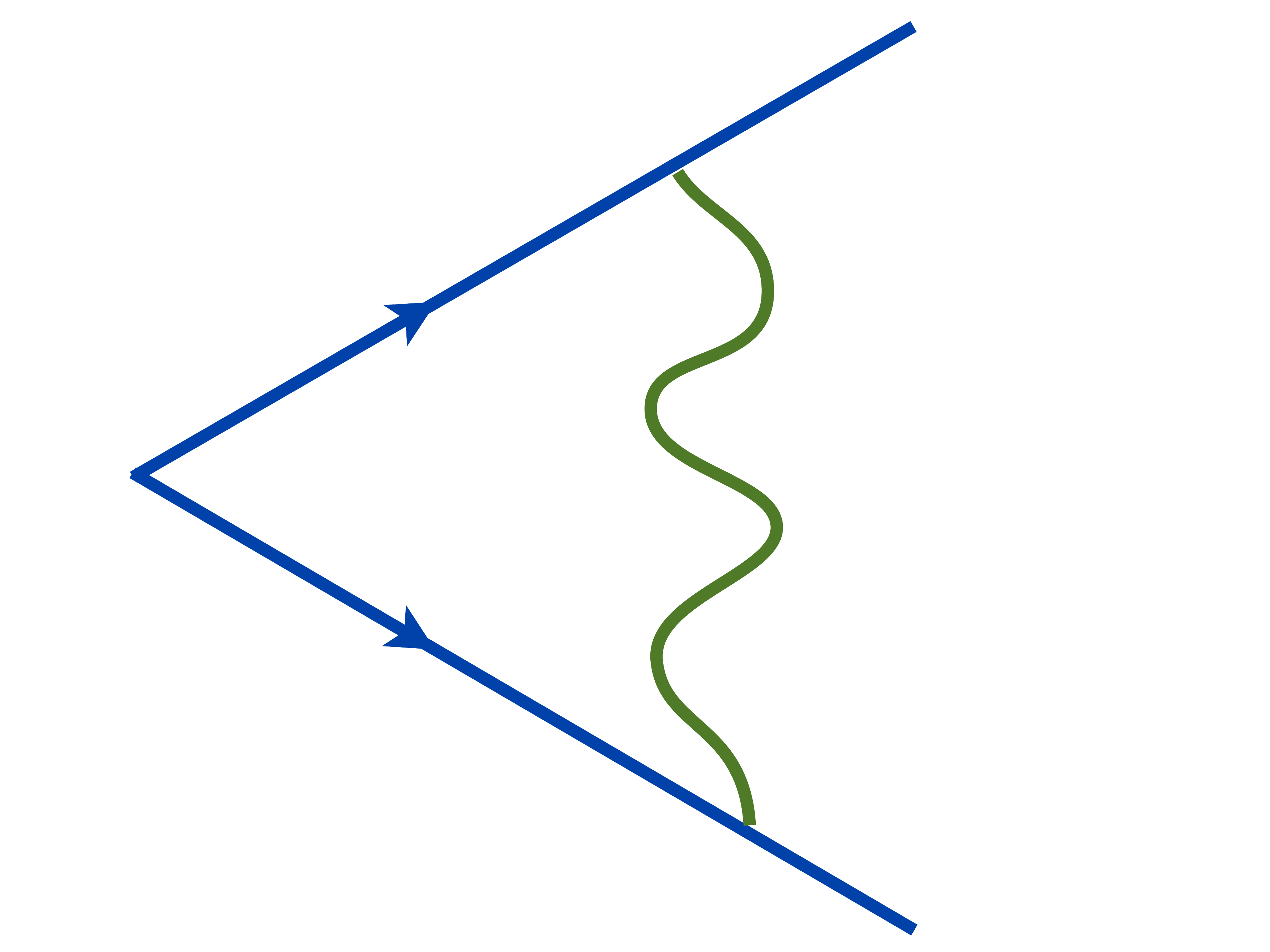}
 \caption{Pairing vertex of two electrons with opposite spin. The wave line represents the `pairing glue':
 the fluctuations of a boson which induces the onset of superconductivity. For the vicinity of the antiferromagnetic
 quantum critical point, the boson is the antiferromagnetic order parameter $\varphi_a$. }
 \label{fig:vertex}
\end{figure}

Let us estimate the strength of this pairing instability by computing the strength of the pairing vertex, $\Lambda$,
as shown in Fig.~\ref{fig:vertex}.
For the standard BCS theory of phonon-mediated superconductivity, this pairing vertex at one-loop order has the form
\beq
\Lambda = 1 + \lambda_{\rm el-ph} \ln \left( \frac{\omega_D}{\omega} \right) \label{bcs}
\eeq
where $\lambda_{\rm el-ph}$ is the electron-phonon coupling, $\omega_D$ is the Debye frequency,
and $\omega \ll E_F$ is the energy of the Cooper pair. This result displays the famous `BCS logarithm', 
which implies that Fermi liquids are ultimately unstable to pairing, and the appearance of superconductivity.
The instability occurs at a frequency/temperature scale at which the vertex correction becomes comparable
to unity, and so the critical temperature $T_c \sim \omega_D \exp ( - 1/\lambda_{\rm el-ph} )$.

Let us now estimate the critical temperature for pairing due to antiferromagnetic fluctuations. Such a computation
can be performed in a fully controlled manner only in the small $U$ limit, where it was found that \cite{raghu}
\beq
\Lambda = 1 + \left( \frac{U}{E_F} \right)^2 \ln \left( \frac{E_F}{\omega} \right), \label{raghu}
\eeq
where $E_F$ is an energy of order the Fermi energy. This result does imply superconductivity, but only
at an exponentially small energy scale in the limit of $U \rightarrow 0$, with $T_c \sim E_F \exp 
\left( - (E_F /U)^2 \right)$.

Finally, let us compute the vertex in Fig.~\ref{fig:vertex} for the theory of the antiferromagnetic critical point described above.
It turns out that the strongest contribution to the pairing vertex does {\em not\/} arise from the non-quasiparticle excitations close to the hot spot (which were considered earlier \cite{fink}): the hot spot is where the pairing glue is the strongest, but the breakdown of the quasiparticle reduces
its efficacy. Instead, the dominant contribution arises from the Fermi lines in its vicinity, where the quasiparticles do survive,
albeit with a small quasiparticle residue, see Eq.~(\ref{Gline}). We can estimate the contribution of these quasiparticles to Fig.~\ref{fig:vertex}, by first computing the Cooper pair propagator
\beq
\int d \Omega \, d k_\perp \,\, G_{\rm Fermi-line} ({\bf k}, \omega+ \Omega) \,\, G_{\rm Fermi-line} ({\bf k}, \Omega) 
\sim \frac{Z^2 (k_\parallel)}{v_F (k_\parallel)} \ln \left( \frac{k_\parallel^2}{\omega} \right)
\eeq
This logarithm is the usual BCS logarithm. Note that it is suppressed by a factor of $Z^2$, indicating that the
logarithm is linked to the integrity of the quasiparticles. From (\ref{kpar}), we see that the prefactor of the BCS
logarithm $\sim \lambda^{-2} k_\parallel$, which is small near the hot spots. However, the vertex in Fig.~\ref{fig:vertex} also involves
a propagator for antiferromagnetic $\varphi_\alpha$ fluctuations, and these contribute a pairing glue 
factor of $\lambda^2/k_\parallel^2$ at the critical point (arising from the gradient term in (\ref{sdw3})). Consequently, we see that the enhancement of the pairing glue at the critical point more than compensates for the
vanishing of the quasiparticle residue, and the remaining integral over $k_\parallel$ is logarithmically divergent.
The final key result is that the correction to the pairing vertex has a $\log^2$ divergence: one logarithm is the BCS logarithm, while the other is a ``quantum critical logarithm' associated with the divergence of spin fluctuations at the critical point.
A careful computation of the $\log^2$ divergence has been carried out \cite{maxsdw1}, and the final result of evaluating
Fig.~\ref{fig:vertex} is
\beq
\Lambda = 1 + \frac{\sin (\theta)}{2 \pi} \ln^2 \left( \frac{E_F}{\omega} \right) \label{log2}
\eeq
There are a number of remarkable features of this key result. First, the $\log^2$
divergence is present for a {\em generic\/} 
antiferromagnetic quantum critical point: no special van Hove singularities are required on the Fermi surface, 
unlike the situation in some early studies \cite{vanhove1,vanhove2}. 
Even more remarkable is the fact that the pre-factor
of the $\log^2$ term in (\ref{log2}) is {\em independent\/} of the Yukawa coupling $\lambda$ in (\ref{sdw3}): the factors of $\lambda$ associated with the pairing glue cancel against the factors of $\lambda$ associated with the renormalization of 
the quasiparticle propagators. Indeed, the prefactor depends only on a geometric feature of the Fermi surface: the angle $\theta$ denoted in Fig.~\ref{fig:fermions}. So we have found an instability towards unconventional superconductivity with a {\em universal strength\/}. The result in (\ref{log2}) implies that $T_c \sim E_F$, which is the promised `mechanism' of
high temperature superconductivity. Values of $T_c$ of this order 
have been discussed earlier \cite{moriya}, but without a universal dimensionless constant characterizing the strength 
of the pairing glue.

Having found this strong instability to pairing, it is now natural to ask if the metal near the onset of antiferromagnetism
has any other instabilities. This question was investigated in Ref.~\cite{maxsdw1}. We only look for
instabilities which are $\log^2$ or stronger in the infrared. This reduces the possibilities greatly, and it was found that there was only
one-additional order parameter with a $\log^2$ enhancement: this was a modulated bond order which is locally an Ising-nematic order; this is reviewed more completely elsewhere \cite{fermiology}. However, crucially, the co-efficient of the
$\log^2$ in the nematic order vertex was smaller than that in (\ref{log2}) by a factor of 3. 
This suggests the dominance
of the instability towards $d$-wave pairing at an energy scale of order $E_F$, of universal strength dependent only
upon geometric features of the Fermi surface. 

\section{The fractionalized Fermi liquid phase}
\label{sec:ffl}

\begin{figure}
\centering
 \includegraphics[width=5.5in]{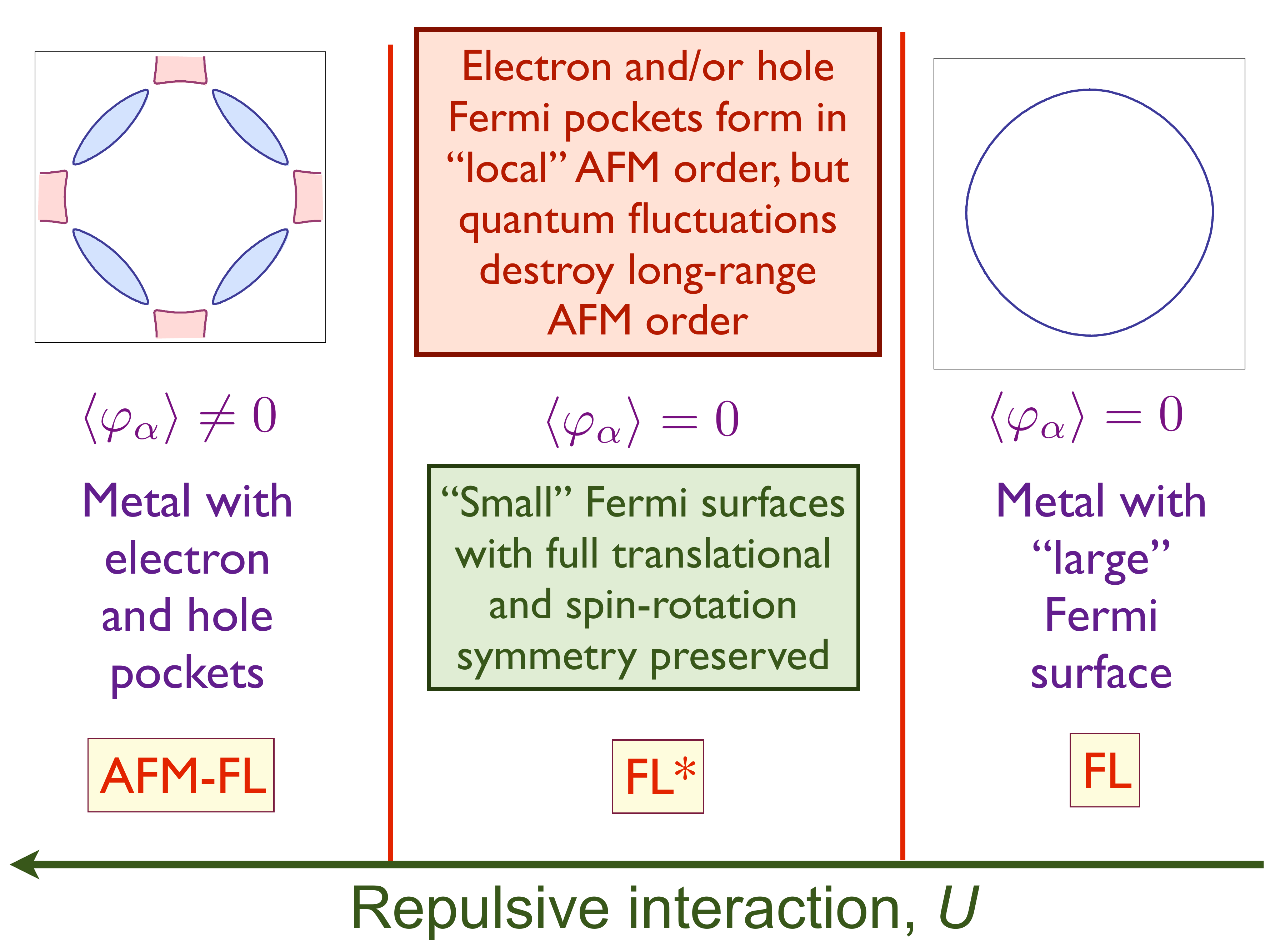}
 \caption{Modification of the phase diagram of Fig.~\ref{fig:sdwphase1}, by the inclusion of an intermediate FL* phase.
 The FL* phase has no antiferromagnetic order ($\langle \varphi_\alpha \rangle = 0$);
 however, at least close to the transition to the metal with antiferromagnetic order, it inherits 
 ``small'' pocket Fermi surfaces from the phase with $\langle \varphi_\alpha \rangle \neq 0$. The FL* phase also has
 additional charge 0 excitations which are similar to those of an insulating spin liquid. The pocket Fermi surfaces
 obey the conventional Luttinger theorem in the AFM-FL phase, but not in the FL* phase.
 }
 \label{fig:sdwphase2}
\end{figure}

We now turn to question (B) from Section~\ref{sec:intro}. Under suitable conditions, can the antiferromagnetic quantum critical point of Fig.~\ref{fig:sdwphase1} be replaced by an exotic intermediate phase ? Here, we describe a route involving
the ``fractionalized Fermi liquid'' (FL*) \cite{ffl1,ffl2,vojtarev}, as shown in Fig.~\ref{fig:sdwphase2}.

In the context of the single-band electronic models considered in Section~\ref{sec:sc} for the cuprates, this FL* phase is best understood near the transition to the AFM-FL. Begin with the AFM-FL, which
has $\langle \varphi_\alpha \rangle \neq 0$ and ``small'' pocket Fermi surfaces obeying the Luttinger relation. 
Now consider quantum fluctuations 
which predominantly involve rotations in the orientation of the local antiferromagnetic order, while maintaining its magnitude.
It seems clear that at least locally, the gap of Fig.~\ref{fig:fermions_gap} 
in the electronic spectrum near the hot spots will be maintained. However,
the traditional picture \cite{lra} is that the fermions will eventually realize the absence of true long-range order, 
and so the hot-spot gap of Fig.~\ref{fig:fermions_gap} 
will fill in at low energies {\em i.e.\/} the gap is only a {\em soft\/}-gap, and there will ultimately be low energy fermionic excitations near the hot spot, and so the Fermi surface will not reconstruct, and the Fermi surface will enclose a ``large'' volume. 
In this case, we revert to the phase diagram discussed in Section~\ref{sec:sc}, of a direct transition from the AFM-FL to 
the ``large'' Fermi surface Fermi liquid without antiferromagnetic long-range order (FL).

However, it has been argued in a series of papers \cite{ss,rkk,kkss,smqx,qi,moon} that there is an alternate possibility: the electronic gap at the hot spots shown in Fig.~\ref{fig:fermions_gap} remains 
a {\em hard\/}-gap at zero temperature, even though true long-range antiferromagnetic order is not present. 
We can roughly understand this gap by transforming to a rotating reference frame oriented along the direction
of the local antiferromagnetic order \cite{smqx}: the resulting fermions will have a hard gap to leading order in the gradients
of the antiferromagnetic order.
The pocket Fermi surfaces survive in a phase {\em without\/} breaking 
of translational or spin rotation symmetry: this is a realization of the FL* phase, in which the small pocket Fermi surfaces do not
obey the Luttinger relation.

The arguments for such a transition build upon a description of the spin fluctuations using `fractionalized'
degrees of freedom \cite{css,senthil1,senthil2,tarun,vojtarev}.
The key idea is that, under suitable conditions, the appropriate bosonic variable for the local 
antiferromagnetic order is not the vector $\varphi_\alpha$,
but a complex bosonic spinor $z_a$: these are related by
\beq
\varphi_{\alpha} = z_a^\ast \sigma^{\alpha}_{ab} z_b . \label{phiz}
\eeq
The $z_a$ spinor can be conveniently used to define a rotating reference frame for the fermions,
oriented along the direction of the local antiferromagnetic order \cite{smqx}.
Note that (\ref{phiz}) is invariant under the U(1) gauge transformation $z_a \rightarrow e^{i \vartheta} z_a$, where the phase $\vartheta$ can have an arbitrary dependence upon spacetime. So $z_a$ is a fractionalized `spinon', which carries unit charge
of an emergent U(1) gauge field; it is, however, neutral under the electromagnetic gauge field. A description 
of the loss of antiferromagnetic order in an AFM-FL phase by
a theory of {\em deconfined\/} $z_a$ spinons leads to an exotic metallic phase without antiferromagnetic order. The $z_a$ spinons
are elementary excitations of this metal, which carry spin $S=1/2$ but are electromagnetically neutral. The electromagnetic charge is 
carried initially by spinless fermions which also carry a charge under the emergent U(1). These fermions have a strong
attractive interaction with the spinons, and so the two bind \cite{rkk,kkss,smqx,qi,moon} 
to form electron-like states carrying spin $S=1/2$ and unit electromagnetic
charge, but which are neutral under the emergent U(1). These bound states fill a Fermi sea, with a small Fermi surface of electron-like
quasiparticles; so we identify this exotic metal as a FL* phase.

This theory of the AFM-FL to FL* transition
has strong orientational fluctuations of the antiferromagnetism, but inhibits magnitude fluctuations by
suppressing topological defects such as hedgehogs, for the case of collinear antiferromagnetism, or $Z_2$ vortices, for the case
of non-collinear magnetism. (Strictly speaking, the hedgehogs are always relevant at long enough scales in a U(1)-FL* 
phase \cite{kkss,qi,ribhu}, but we will ignore this here, assuming the crossover to confinement happens at temperatures
lower than those of interest to us.)
The suppression of defects turns out to be sufficient to allow the hot spot gap of Fig.~\ref{fig:fermions_gap} to survive.

Recently, a more direct description of the AFM-FL to FL* transition has been achieved \cite{mpss}. 
This approach avoids the intermediate regime with the spinless fermion states noted above, and deals directly
with the electron-like bound states using 
the `spinon-dopon' formulation of Ribeiro and Wen \cite{RW1}.

The FL* phase is a metal which breaks no symmetries, but differs from the conventional Fermi liquid (FL) 
in two crucial ways: 
\begin{itemize}
\item The FL* phase has gapless $S=1/2$, charge $e$ quasiparticle excitations, just like a FL, but
the number of these excitations is different. In a FL, the gapless quasiparticles lie on a Fermi surface which encloses a 
``large'' volume equal to the total density of electrons: this is the familiar Luttinger theorem. In contrast, in a FL* phase,
the Fermi surface of electron-like excitations has a volume which differs from the total density by one electron 
per unit cell: this leads
to the ``small'' pocket Fermi surfaces, which now violate the conventional Luttinger relation. 
\item
The second important difference is that in a FL the Fermi surface quasiparticles are the {\em only\/} low energy excitations, while the FL* phase also has neutral $S=1/2$ spinon and associated gauge excitations.
\end{itemize}

Indeed, these two distinctions between the FL and FL* phase are intimately linked. The link is provided by 
Oshikawa's non-perturbative proof of the Luttinger theorem for the Fermi liquid \cite{oshikawa}. 
Oshikawa used 
a topological argument analogous to Laughlin's argument for the quantization of the Hall conductance. A key ingredient in his proof was the assumption that the only low energy excitations of the quantum state 
are the quasiparticles at the Fermi surface, which is true in the FL phase. However, it was subsequently noted \cite{ffl2}
that this assumption provided an escape hatch. Insulating spin liquids invariably have low energy global topological excitations of the type accounted for by Oshikawa's argument. The FL* phase inherits these global topological excitations from the spin liquid: in our analysis above, these excitations are associated with the emergent gauge field present in the theory of the deconfined
$z_a$ spinons. Using Oshikawa's method, it was then shown \cite{ffl2} that the global topological excitations of the FL* allowed
violation of the conventional Luttinger count on the volume enclosed by the Fermi surface. Instead, this modification of the Oshikawa argument leads to a FL* phase with ``small'' pocket Fermi surfaces, enclosing the same total volume as those in the AFM-FL.
We reiterate that the AFM-FL phase does not posses these topological excitations, but its small Fermi surfaces do obey the conventional Luttinger count because of the doubling of the unit cell by the antiferromagnetic order. 

One significant consequence of these arguments is that now the Fermi surface volume in the FL* phase can
be viewed as a direct  
experimental signature of the topological order of the spin liquid. In insulators, the topological order has
so far evaded experimental detection; remarkably, in metals its detection requires only measurement of the
Fermi surface volume by photoemission, and so is straightforward.

\begin{figure}
\centering
 \includegraphics[width=3.8in]{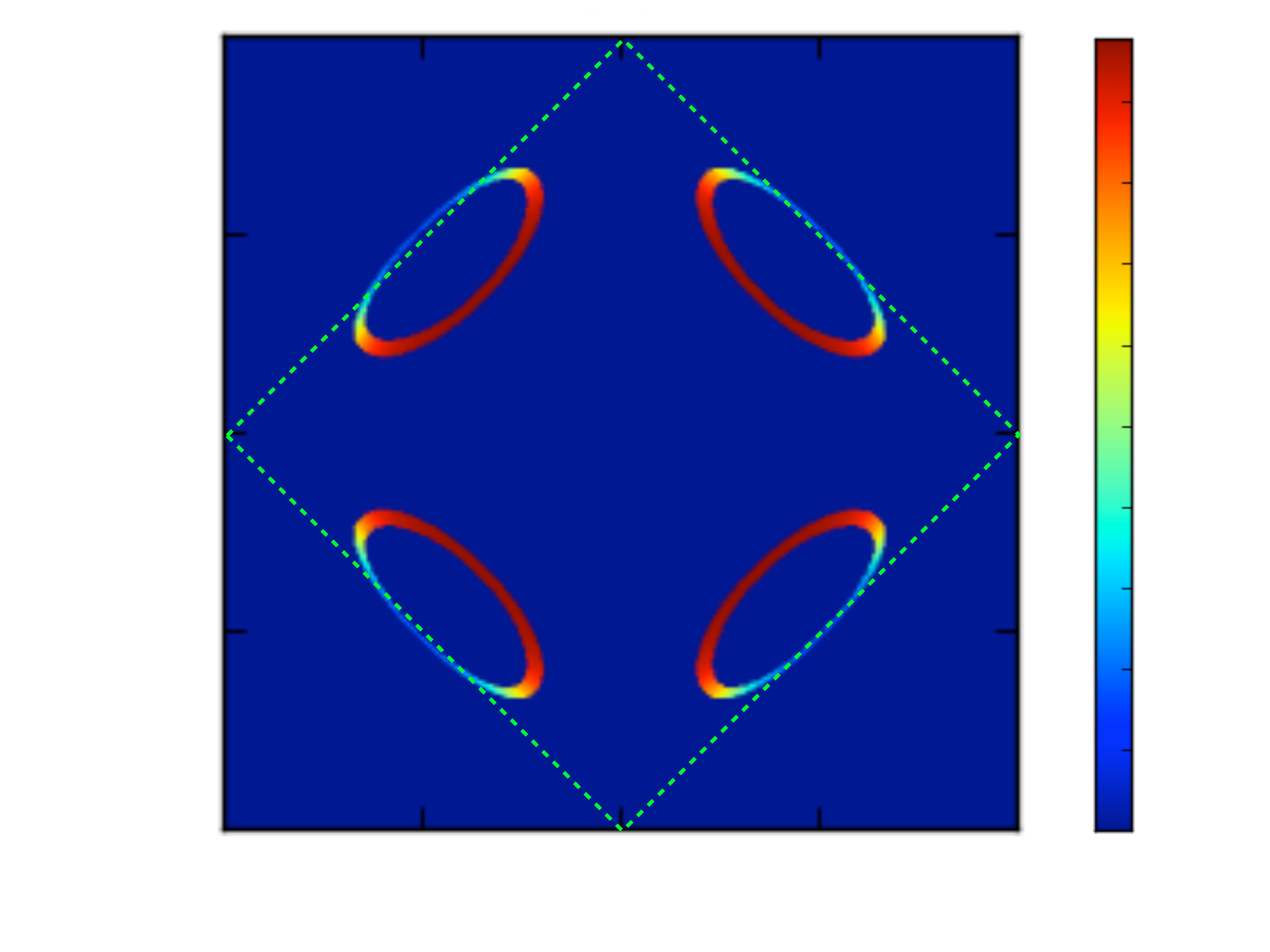}
 \caption{Fermi surfaces in the FL* phase, computed in the phenomenological model of Ref.~\cite{qi}. The color scale
 represents the quasiparticle residue on the Fermi surface. Note that the hole pocket Fermi surfaces are not centered on the magnetic Brillouin zone boundary, or otherwise sensitive to it. The volume enclosed by the pocket Fermi surfaces is the same
 as that in the AFM-FL phase, but the conventional Luttinger relation is violated only in the FL* phase.}
 \label{fig:qi}
\end{figure}

What about the {\em shape\/} of the pocket Fermi surfaces in the FL* phase ? In the AFM-FL, these pockets were created by Bragg reflection of the Fermi surface across the magnetic Brillouin zone boundary; consequently, the pockets are always centered on the magnetic Brillouin zone boundary. For the FL* phase, this question was addressed using a phenomenological effective
field theory in Ref.~\cite{qi}, and the results are shown in Fig.~\ref{fig:qi}.
Now there is full symmetry of the square lattice, and the magnetic Brillouin zone boundary plays no special role. Consequently, the hole pocket Fermi surfaces are centered at a generic point in momentum space, which generally does not lie on the magnetic Brilluoin zone.

The FL* phase described here is a candidate for the `pseudogap regime' of the hole-doped cuprates. The gapping of the $z_a$ spinons, and of a large portion of the Fermi surface, can account for the reduction of the spin susceptibility.
The hole pocket spectrum in Fig.~\ref{fig:qi} has similarities to photoemission observations, in particular to Ref.~\cite{yang}.
At low temperatures, the FL* phase may be unstable to confinement transitions similar to those found in insulating spin 
liquids \cite{rs1,ribhu}, and this would lead to translational symmetry breaking due to valence-bond ordering, and this may be connected to scanning tunnelling microscopy observations \cite{kohsaka}.
NMR measurements \cite{julien} 
on YBa$_2$Cu$_3$O$_y$ have not observed antiferromagnetic order at fields upto 30 Tesla, but do see indications of charge ordering.
The angle dependence of quantum oscillations
in YBa$_2$Cu$_3$O$_{6.59}$ has been argued \cite{ramshaw} to imply the absence of spin-density wave ordering.

In the above experimental application, the main role of the FL* physics is to provide a simple route to obtaining pocket Fermi surfaces
without antiferromagnetic order. Should charge/valence-bond order appear at the lowest scales, and the unit cell increases in size,
the Fermi volumes of the FL* become compatible with the Luttinger volume. Nevertheless, the FL* phase can remain distinct from
a Fermi liquid due to the presence of spinon and gauge excitations. A true Fermi liquid is obtained only if a confinement transition
eliminates these extraneous excitations. 

\subsection{The Kondo lattice and the heavy fermion materials}
\label{sec:kondo}

We now give a different perspective on the FL* phase, appropriate for application to the heavy fermion materials.
Rather than working with analogs of the single-band Hubbard model used so far for the cuprates, we formulate the theory
in terms of phases of the Kondo lattice model. We will find phases with the same qualitative low energy characteristics, and so will identify them with the same labels. However, the short-distance physical interpretation will be different, and this will give
additional insight into the physics of these phases. In particular, we will find that the FL* phase appears more naturally, and has a simple physical interpretation. 

\begin{figure}
\centering
 \includegraphics[width=5in]{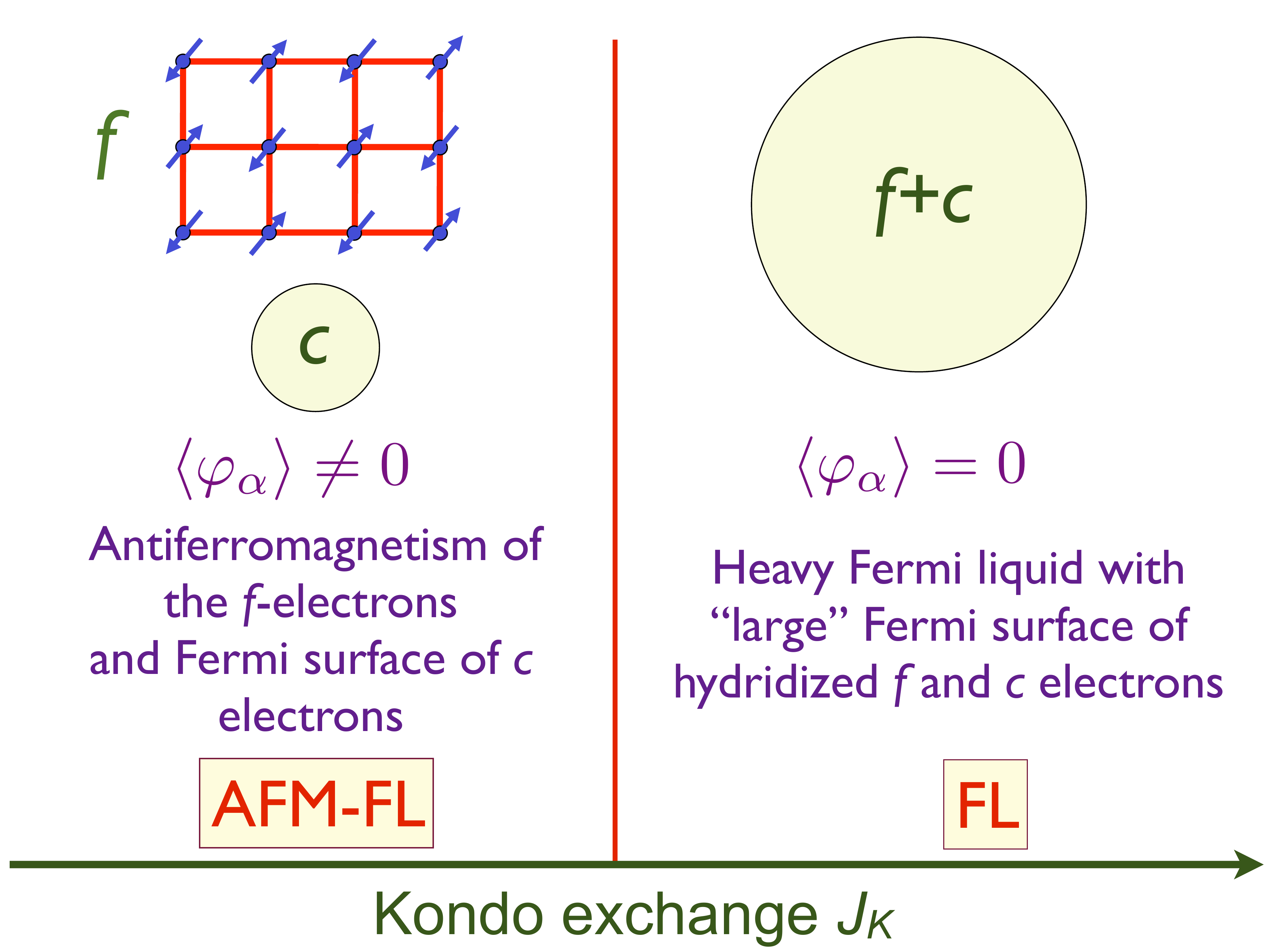}
 \caption{Fermi liquid phases of the Kondo lattice model of $f$ local moments and $c$ conduction electrons. 
 These are qualitatively identical to the phases of the single-band
 Hubbard model in Fig.~\ref{fig:sdwphase1}. The Fermi surfaces of both phases obey the conventional Luttinger theorem.}
 \label{fig:kondophase1}
\end{figure}

The Kondo lattice model is described in terms of two bands of electrons: the localized $f$ electrons, and the itinerant conduction electrons, $c$. The $f$ electrons interact with each other via direct exchange interactions labeled $J_H$, and with the $c$ electrons via the Kondo exchange $J_K$. As a function of the ratio $J_K/J_H$, there are two basic Fermi liquid phases, which are shown in Fig.~\ref{fig:kondophase1}.
This phase diagram is the analog of Fig.~\ref{fig:sdwphase1} for the single band model.

For large $J_K/J_H$, we can initially treat the $f$ moments as independent. Each $f$ moment is Kondo-screened by
the conduction electrons, and this is described in Wilson's renormalization group treatment as a flow of $J_K \rightarrow \infty$.
For the lattice model, this Kondo screening leads to the well-studied heavy Fermi liquid state in which there is a ``large''
Fermi surface enclosing a volume counting the density of both the $f$ and $c$ electrons. Apart from its two-band nature,
and the large quasiparticle mass, this phase is not fundamentally distinct from the FL state of the 
single-band model in Fig.~\ref{fig:sdwphase1},
and so we have identified it accordingly in Fig.~\ref{fig:kondophase1}. For both models, the FL state is adiabatically connected
to the trivial Fermi liquid state of non-interacting electrons.

In contrast, for large $J_H/J_K$, the exchange between the $f$ electrons can lead to antiferromagnetic order. If this order is strong enough, we can treat the $f$ moments as static, and then the $c$ electrons are free to form
their own Fermi liquid. This Fermi surface of $c$ electrons is small, but the Luttinger relation is obeyed because of the 
doubling of the unit cell by the antiferromagnetic order. Again, the resulting AFM-FL state is qualitatively identical to that
of the single band model in Fig.~\ref{fig:sdwphase1}, and so has been identified by the same symbol in 
Fig.~\ref{fig:kondophase1}. There may be additional distinctions within the AFM-FL state involving changes in the shape of the Fermi surface while preserving its volume: we are ignoring these here.

\begin{figure}
\centering
 \includegraphics[width=5.5in]{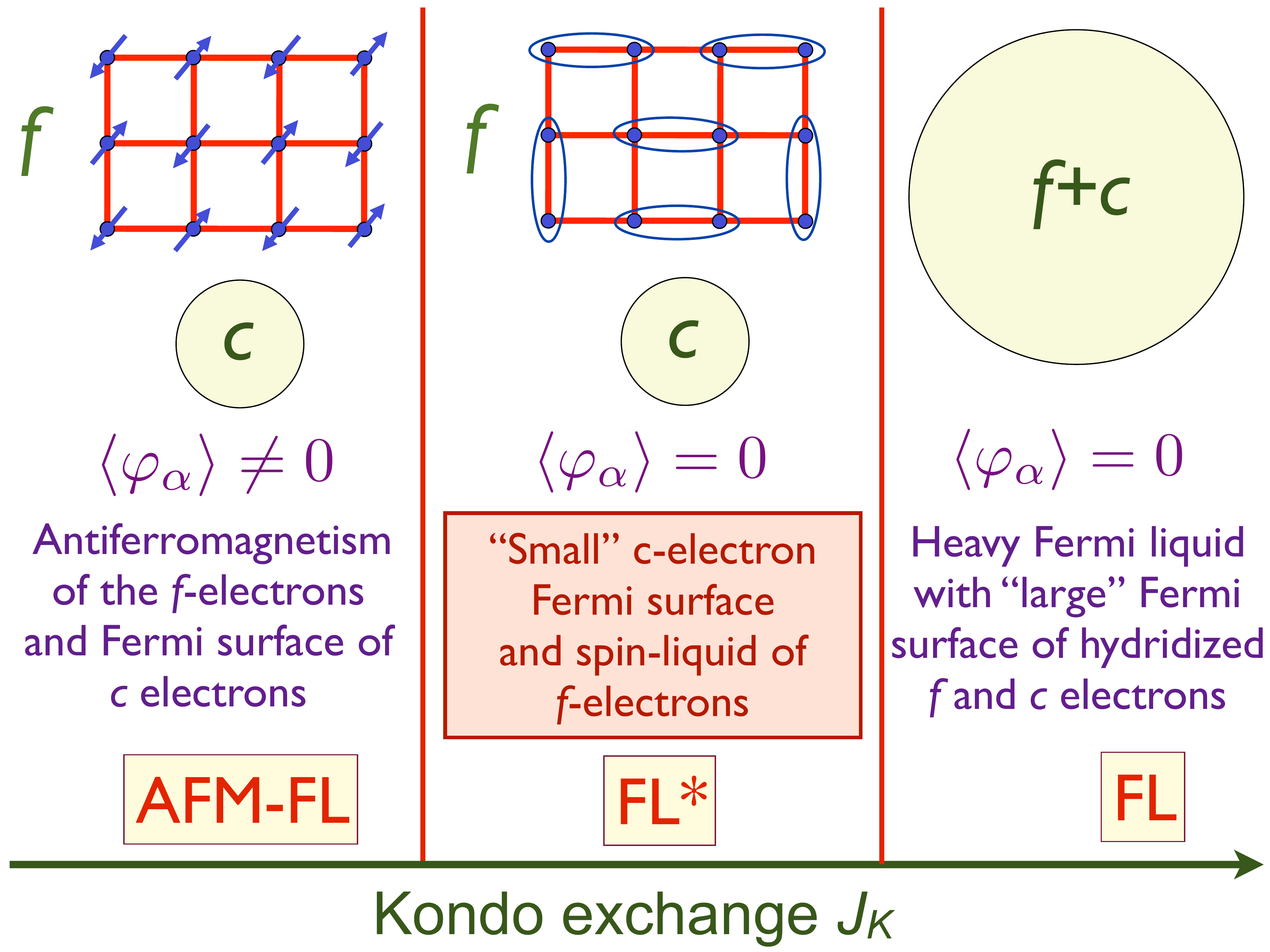}
 \caption{Extension of Fig.~\ref{fig:kondophase1} for the case where the $f$-$f$ exchange is frustrated and can induce
 a spin liquid of the $f$ moments. Notice the similarity to the one-band phase diagram in Fig.~\ref{fig:sdwphase2}. The FL* is now understood simply by adiabatic continuity to a state in which a spin-liquid of $f$ moments is decoupled from
 a small Fermi surface of $c$ electrons; this small Fermi surface does not obey the conventional Luttinger relation.
 }
 \label{fig:kondophase2}
\end{figure}

However, just as in Fig.~\ref{fig:sdwphase2} for the single-band model, there is also the possibility here of an 
intermediate FL* phase \cite{ffl1,ffl2,bgg}, as shown in Fig.~\ref{fig:kondophase2}.
Consider the situation where $J_H/J_K$ is large, so we can initially ignore the Kondo exchange. Also, choose the $J_H$
so that the $f$-$f$ exchange is frustrated, leading to a spin liquid ground state for the $f$ electrons: 
this could happen {\em e.g.\/} if the
$f$ moments reside on a triangular lattice. Now let us examine the influence of $J_K$. Unlike the independent moment
limit usually studied in the Kondo model, now we don't have a flow at low energies to $J_K \rightarrow \infty$: the $f$ spin liquid lifts the two-fold degeneracy of each independent spin, and this quenches the renormalization group flow of $J_K$. Consequently, the resulting state of the Kondo lattice model is now similar to 
the $J_K \rightarrow 0$ state, rather than to the $J_K \rightarrow \infty$ state. This is the FL* state of Fig.~\ref{fig:kondophase2}, with a
small Fermi surface of $c$ electrons;
because all symmetries are preserved
and there is no doubling of the unit cell, the conventional Luttinger relation is violated.
The Fermi surface of this FL* phase is associated with the band structure of the $c$ electrons alone, in contrast to its
dependence upon local antiferromagnetism in the single band model discussed earlier. 

\begin{figure}
\centering
 \includegraphics[width=5.5in]{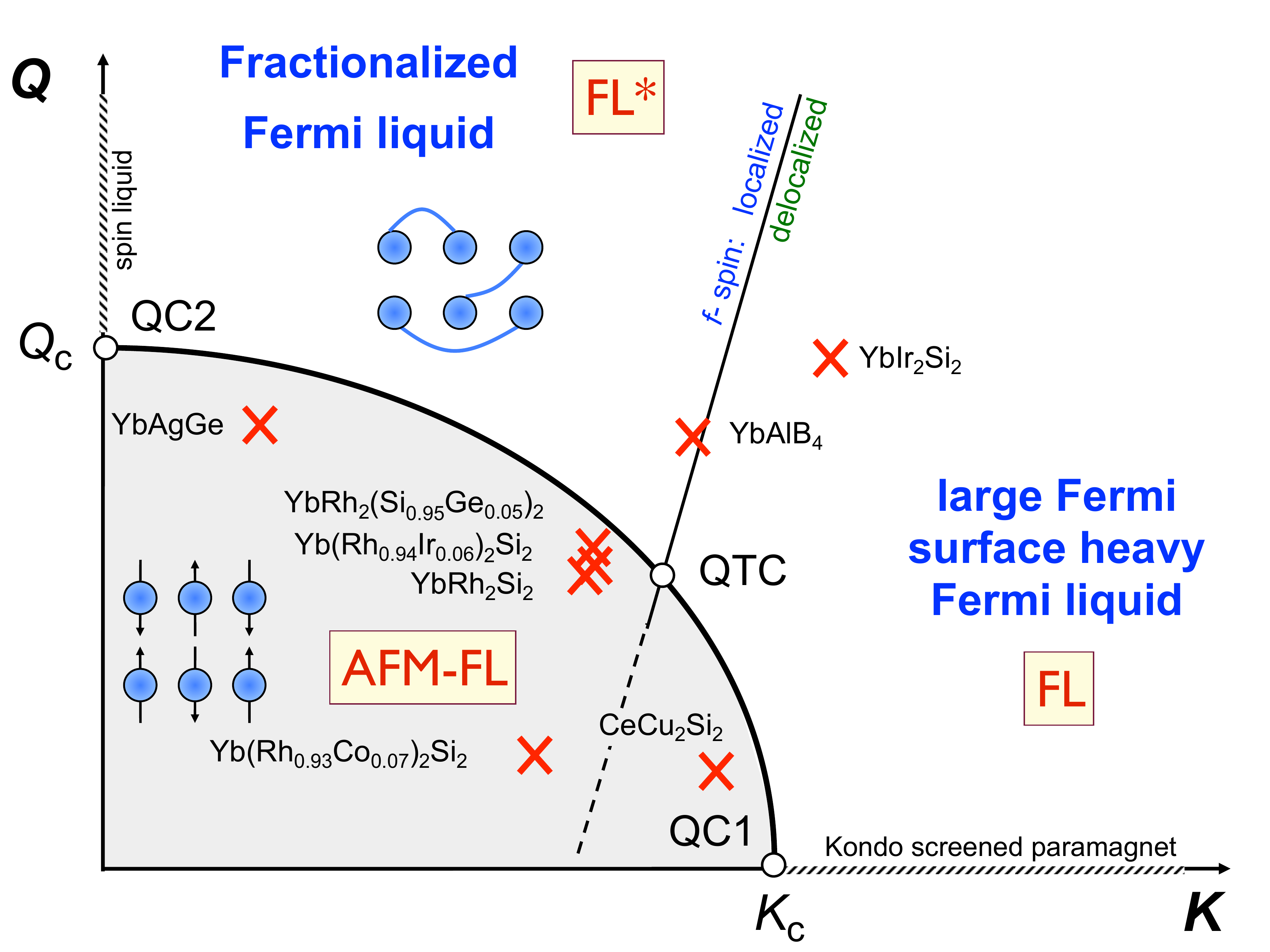}
 \caption{Phase diagram of Custers {\em et al.} \cite{steglich2} (with permission; see also Ref.~\cite{silke1}). The labels of the phases have been changed to correspond to those in Figs.~\ref{fig:kondophase1},\ref{fig:kondophase2}, and an extra point has been
 added for CeCu$_2$Si$_2$. Here $K$ represents a Kondo exchange, and $Q$ is a measure of the frustration in the
 $f$-$f$ exchange. At small values of $Q$, the phase diagram as a function of increasing $K$ is as in Fig.~\ref{fig:kondophase1};
 and at large values of $Q$, the phase diagram as a function of increasing $K$ is as in Fig.~\ref{fig:kondophase2}.}
 \label{fig:coleman}
\end{figure}

The combination of Figs.~\ref{fig:sdwphase1} and Fig.~\ref{fig:sdwphase2} is a useful framework for understanding the physics
of a wide variety of heavy fermion compounds \cite{ffl1,ffl2,bgg,si}. Initial evidence for an intermediate FL* state between the well-studied
AFM-FL and FL phase appeared in the field-tuned studies of YbAgGe by Bud'ko {\em et al.} \cite{budko}.
More recently, the extensive studies of YbRh$_2$Si$_2$ \cite{steglich1,steglich2} are so far consistent with a FL* interpretation. 
We reproduce in Fig.~\ref{fig:coleman} a phase diagram of Custers {\em et al.} \cite{steglich2,silke1}, which
combines our Figs.~\ref{fig:kondophase1} and~\ref{fig:kondophase2} with additional experimental information. 

\subsection{Discussion}
\label{sec:dis}

We have given two complementary descriptions of the FL* phase above. 

First, we approached the FL* phase from the AFM-FL
phase. In this case, the antiferromagnetic order, and its subsequent `quantum disordering' was described most conveniently
by a theory of bosonic spinons. Consequently, the resulting FL* state had bosonic spinon excitations, reflecting the
nature of the underlying spin liquid.

Our second treatment of the FL* state used a Kondo lattice model. This approach most conveniently describes the FL* to FL
transition \cite{ffl2}, using an underlying spin liquid with fermionic spinons. We note that a fermionic spinon
approach has been used recently \cite{RW1,RW2,RW3,weng} to describe the under-doped cuprates 
as a ``Luttinger-volume violating Fermi liquid'' (LvvFL):
the LvvFL state is qualitatively the same as the FL* state.

Therefore, we don't have a single theory which can fully describe all the phases of Figs.~\ref{fig:sdwphase2} and~\ref{fig:kondophase2}, and follow the evolution of the Fermi surface across the two (or more) quantum phase transitions.
At the very least, we need a description of the transmutation of the neutral spinon excitations of the FL* phase from
fermions to bosons. 
Finding such a theory remains an important problem for future theoretical research. 

Finally, we note that the FL* phase appears naturally as the correlated metallic state 
in a large number of recent studies of compressible states by holographic methods \cite{ssffl}.

\subsection*{Acknowledgements}
This research was supported by the National Science Foundation under grant DMR-1103860, 
and by a MURI grant from AFOSR. MM was also supported by the NSF
under Grant No. NSF PHY11-25915.

\end{document}